\documentclass[manuscript]{aastex}

\shorttitle{Rising Motion of Solar Magnetic Fields}
\shortauthors{Toriumi et al.}

\begin{document}

%% LaTeX will automatically break titles if they run longer than
%% one line. However, you may use \\ to force a line break if
%% you desire.

\title{Probing the Shallow Convection Zone:
  Rising Motion of Subsurface Magnetic Fields
  in the Solar Active Region}

%% Use \author, \affil, and the \and command to format
%% author and affiliation information.
%% Note that \email has replaced the old \authoremail command
%% from AASTeX v4.0. You can use \email to mark an email address
%% anywhere in the paper, not just in the front matter.
%% As in the title, use \\ to force line breaks.

\author{Shin Toriumi$^{1}$,
  Stathis Ilonidis$^{2}$,
  Takashi Sekii$^{3}$,
  and Takaaki Yokoyama$^{1}$}
\affil{$^{1}$Department of Earth and Planetary Science,
  University of Tokyo, 7-3-1 Hongo, Bunkyo-ku, Tokyo 113-0033, Japan}
\email{toriumi@eps.s.u-tokyo.ac.jp}
\affil{$^{2}$W.W. Hansen Experimental Physics Laboratory,
  Stanford University, Stanford, CA 94305-4085}
\affil{$^{3}$National Astronomical Observatory of Japan,
  2-21-1 Osawa, Mitaka, Tokyo 181-8588, Japan}

%% Notice that each of these authors has alternate affiliations, which
%% are identified by the \altaffilmark after each name.  Specify alternate
%% affiliation information with \altaffiltext, with one command per each
%% affiliation.

%% Mark off your abstract in the ``abstract'' environment. In the manuscript
%% style, abstract will output a Received/Accepted line after the
%% title and affiliation information. No date will appear since the author
%% does not have this information. The dates will be filled in by the
%% editorial office after submission.

\begin{abstract}
  In this Letter
  we present a seismological detection
  of a rising motion
  of magnetic flux
  in the shallow convection zone
  of the Sun,
  and show estimates
  of the emerging speed
  and its decelerating nature.
  In order to evaluate
  the speed of
  subsurface flux
  that creates an active region,
  we apply six Fourier filters
  to the Doppler data
  of NOAA AR 10488,
  observed with SOHO/MDI,
  to detect the reduction
  of acoustic power
  at six different depths
  from $-15$ to $-2\ {\rm Mm}$.
  All the filtered acoustic powers
  show reductions,
  up to 2 hours
  before the magnetic flux
  first appears
  at the visible surface.
  The start times of these reductions
  show a rising trend
  with a gradual deceleration.
  The obtained velocity
  is first
  several ${\rm km\ s}^{-1}$
  in a depth range of 15--10 Mm,
  then $\sim 1.5\ {\rm km\ s}^{-1}$
  at 10--5 Mm,
  finally $\sim 0.5\ {\rm km\ s}^{-1}$
  at 5--2 Mm.
  If we assume that
  the power reduction
  is actually caused
  by the magnetic field,
  the velocity of order of $1\ {\rm km\ s}^{-1}$
  is well in accordance
  with previous observations and numerical studies.
  Moreover,
  the gradual deceleration
  strongly supports
  the theoretical model
  that the emerging flux
  slows down
  in the uppermost convection zone
  before it expands into the atmosphere
  to build an active region.
\end{abstract}

%% Keywords should appear after the \end{abstract} command. The uncommented
%% example has been keyed in ApJ style. See the instructions to authors
%% for the journal to which you are submitting your paper to determine
%% what keyword punctuation is appropriate.


%% From the front matter, we move on to the body of the paper.
%% In the first two sections, notice the use of the natbib \citep
%% and \citet commands to identify citations.  The citations are
%% tied to the reference list via symbolic KEYs. The KEY corresponds
%% to the KEY in the \bibitem in the reference list below. We have
%% chosen the first three characters of the first author's name plus
%% the last two numeral of the year of publication as our KEY for
%% each reference.

%% Authors who wish to have the most important objects in their paper
%% linked in the electronic edition to a data center may do so by tagging
%% their objects with \objectname{} or \object{}.  Each macro takes the
%% object name as its required argument. The optional, square-bracket 
%% argument should be used in cases where the data center identification
%% differs from what is to be printed in the paper.  The text appearing 
%% in curly braces is what will appear in print in the published paper. 
%% If the object name is recognized by the data centers, it will be linked
%% in the electronic edition to the object data available at the data centers  
%%
%% Note that for sources with brackets in their names, e.g. [WEG2004] 14h-090,
%% the brackets must be escaped with backslashes when used in the first
%% square-bracket argument, for instance, \object[\[WEG2004\] 14h-090]{90}).
%%  Otherwise, LaTeX will issue an error. 

\section{Introduction
  \label{sec:introduction}}

It has long been believed that
solar active regions (ARs)
are the consequence
of rising magnetic fields
from the deep convection zone,
i.e., flux emergence
\citep{par55}.
Theoretically,
the buoyancy
of the magnetic flux
initially accelerates
its ascent
\citep{fan09}.
As it approaches the surface,
however,
accumulated plasma between the rising flux
and the isothermally stratified photospheric layer
decelerates the ascent of the flux
\citep{tor10,tor11,tor12}.
However,
we cannot investigate the physical state
(e.g. rising speed)
of the subsurface magnetic flux
from direct optical observations.

One possible way
to overcome this problem
is helioseismology,
although previously
it was thought to be difficult
to detect any significant seismic signatures
associated with the emerging flux
before it appears at the surface
because of the fast emergence
and low signal-to-noise (S/N) ratio
\citep{kos09}.
Recent observation by \citet{ilo11}, however,
detected strong acoustic travel-time anomalies
1--2 days before the photospheric flux
attains its peak flux emergence rate.
They estimated
the flux rising speed
from $-65\ {\rm Mm}$
of the convection zone
to the surface
to be $0.3$--$0.6\ {\rm km\ s}^{-1}$.
\citet{har11} focused on
the surficial acoustic oscillation power
(time-averaged squared velocity)
and found that a reduction
in acoustic power
in the frequency range of
$3$--$4\ {\rm mHz}$
can be seen about 1 hr
before the start of the flux appearance.
Their interpretation was that
the acoustic power
is reduced by the subsurface magnetic field.

In this Letter,
we present the first detection
of the ``rising motion''
corresponding to the emerging magnetic flux,
by using
acoustic power measurement
at the surface.
The idea behind this study
is as follows:
It is possible to apply
a Fourier filter to the Doppler data,
to extract acoustic waves
penetrating to a particular range of depths.
A scheme similar to deep focusing
\citep{duv03},
with annuli set up
around the surface points
above the targets,
can also be applied
to integrate the signal
from that depth range.
The acoustic power of
such a filtered wavefield,
then, must be influenced
by the acoustic power
in this depth range.
That is,
if the acoustic power
is reduced
in a certain region
in the solar interior
by a power-reducing
agent
such as magnetic field
\footnote{
Here we assume that
waves may be locally damped,
or scattered off
their original paths,
by such an agent,
resulting in surface power reduction.
Therefore, for simplicity,
we use the term ``power-reducing agent.''
},
the acoustic power
observed in the surface regions
which are connected
to this region through rays
corresponding to the observed wave components may
also be reduced.
Therefore,
in this study,
to the Doppler data
in an emerging AR,
we apply six different filters
that have 
primary sensitivities
to six different depths
according to ray theory,
and see the temporal evolutions
of acoustic power
that may correspond
to those depths.
If the power reduction starts
from the deeper layer,
we can speculate that
the power-reducing agent
is rising in the interior.
In this analysis,
we focus on
the uppermost convection zone
just before the flux emergence
at the visible surface.

\section{Data Analysis
  \label{sec:analysis}}

\begin{figure}
  \includegraphics[scale=1.,clip]{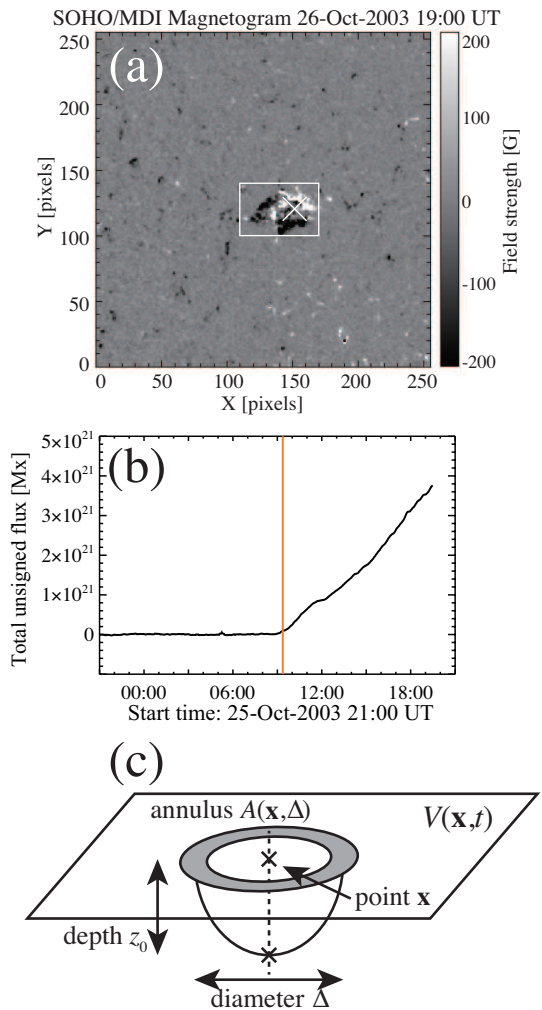}
  \caption{(a) Magnetogram of NOAA AR 10488.
    The field of view is
    the same as Doppler data
    used in this analysis.
    The center of X
    indicates the emergence location
    $\mbox{\boldmath $x$}_{1}$.
    (b) Time-evolution of the total unsigned flux
    calculated in the white window in Panel (a).
    The vertical line shows the start of the flux emergence
    measured by the method in \citet{tor12b}
    (c) Schematic illustration
    of calculating the acoustic power
    from the filtered Doppler velocity data.
    See text for details.
  }
  \label{fig:magneto}
\end{figure}

Doppler observations
of NOAA AR 10488
(Figure \ref{fig:magneto}(a))
from Michelson Doppler Imager
\citep[MDI;][]{sch95}
on board the {\it Solar and Heliospheric Observatory}
({\it SOHO})
are used in this work.
Time-evolution of
the total unsigned magnetic flux
in this AR
is plotted
as Figure \ref{fig:magneto}(b).
The tracked Doppler dataset
(AR data)
has a cadence of 1 min
with a duration of 24 hr
from 19:30 UT, 2003 October 25
(about 14 hr before the flux appearance)
and a size of $256\times 256$ pixels
with a pixel size of 0.12 degree
in the heliographical coordinates.
For comparison
we also use the Doppler data
at the same disk location
but 17 days later
when there was no active region
(QS data).
Here it should be noted that
each dataset has some temporal gaps
(periods of no observation)
in the whole 24 hr data
\footnote{
  AR data: 19:53--20:01 UT
  on October 25th
  and 05:11--05:19 UT on 26th.
  QS data: 02:57--03:05 UT
  and 13:12--13:26 UT
  on November 12th.
}.
According to \citet{har11},
the gaps may cause
significant variations
in the power.
Therefore,
in the 24 hr data,
we discuss the power evolution
only between 06:00--12:00 UT,
during which there is no temporal gap,
to avoid the effect of the gap.

First we eliminate the signal
below $1.5\ {\rm mHz}$ and above $5.5\ {\rm mHz}$
from Fourier-transformed data
both in AR and QS,
using a box-car filter in the frequency domain
with a hyperbolic-tangent roll-off.
To eliminate the contribution of the f-mode,
we also apply a high-pass filter
with a Gaussian roll-off.
Then we consider
two types of Fourier filters,
one of which is
applied to the both datasets.
The first type of filters is
a series of phase-speed filters
constructed based on the parameters used in
\citet{zha12}.
The second type is
ridge filters
which extract the power
of p1, p2, and p3-modes.
The ridge filters are constructed
by approximating
the location of each ridge
and have a box-car shape
in the frequency domain
with hyperbolic-tangent roll-offs.
The properties of the filters
are listed in Table \ref{tab:filter}.

To determine the depth to which each filter is most sensitive,
we examined how the filtered power is distributed over
the phase speed $V_{\rm ph}=\omega/k_{\rm h}$, by constructing
a histogram indicating the power, for each phase-speed bin,
summed over the corresponding $k_{\rm h}$--$\omega$ bins. We
identify the mode of this distribution as the effective
phase speed, and then find the target depth $z_{0}$ as the
asymptotic inner turning point corresponding to this
phase speed, using the model S of
\citet{chr96}.
The target depth $z_{0}$
and its error,
which is
determined from the width
of the histogram,
are shown in Table \ref{tab:filter}.
Here, $\omega$ and $k_{\rm h}$
are the angular frequency
and horizontal wavenumber
of a sound wave,
respectively.

Finally, we produce
the acoustic power maps
from the filtered Doppler velocity data.
In order to increase
the S/N ratio,
the acoustic power
at the point $\mbox{\boldmath $x$}$
at the time $t$,
$P($\mbox{\boldmath $x$}$,t)$,
is given as the squared velocity
averaged over an annulus
of a diameter
of one travel distance $\Delta$
centered at $\mbox{\boldmath $x$}$,
${\cal A}($\mbox{\boldmath $x$}$,\Delta)$:
\begin{eqnarray}
  P (\mbox{\boldmath $x$},t)
  = \frac{
    \displaystyle
    \sum_{\mbox{\boldmath $x$}' \in {\cal A}}
    \Bigl[ V (\mbox{\boldmath $x$}',t) \Bigr]^{2}
    }{
      \displaystyle
      \sum_{\mbox{\boldmath $x$}' \in {\cal A}}
      \delta S (\mbox{\boldmath $x$}')
    },
\end{eqnarray}
where $V($\mbox{\boldmath $x$}$,t)$
is the filtered velocity
and $\delta S($\mbox{\boldmath $x$}$)$
is the area of a pixel element.
The schematic illustration
of the power calculation
is shown as Figure \ref{fig:magneto}(c).
If the acoustic wave is affected
by the power-reducing agent
at the bottom of the ray path,
the observed acoustic power
in the annulus
will be reduced.
For a higher S/N,
the thickness of the annulus
is made twice as broad as
shown in Table \ref{tab:filter}.
The obtained maps show
the temporal and two-dimensional evolution
of acoustic powers
at six different depths.

\section{Results
  \label{sec:results}}

\begin{figure}
  \includegraphics[scale=1.,clip]{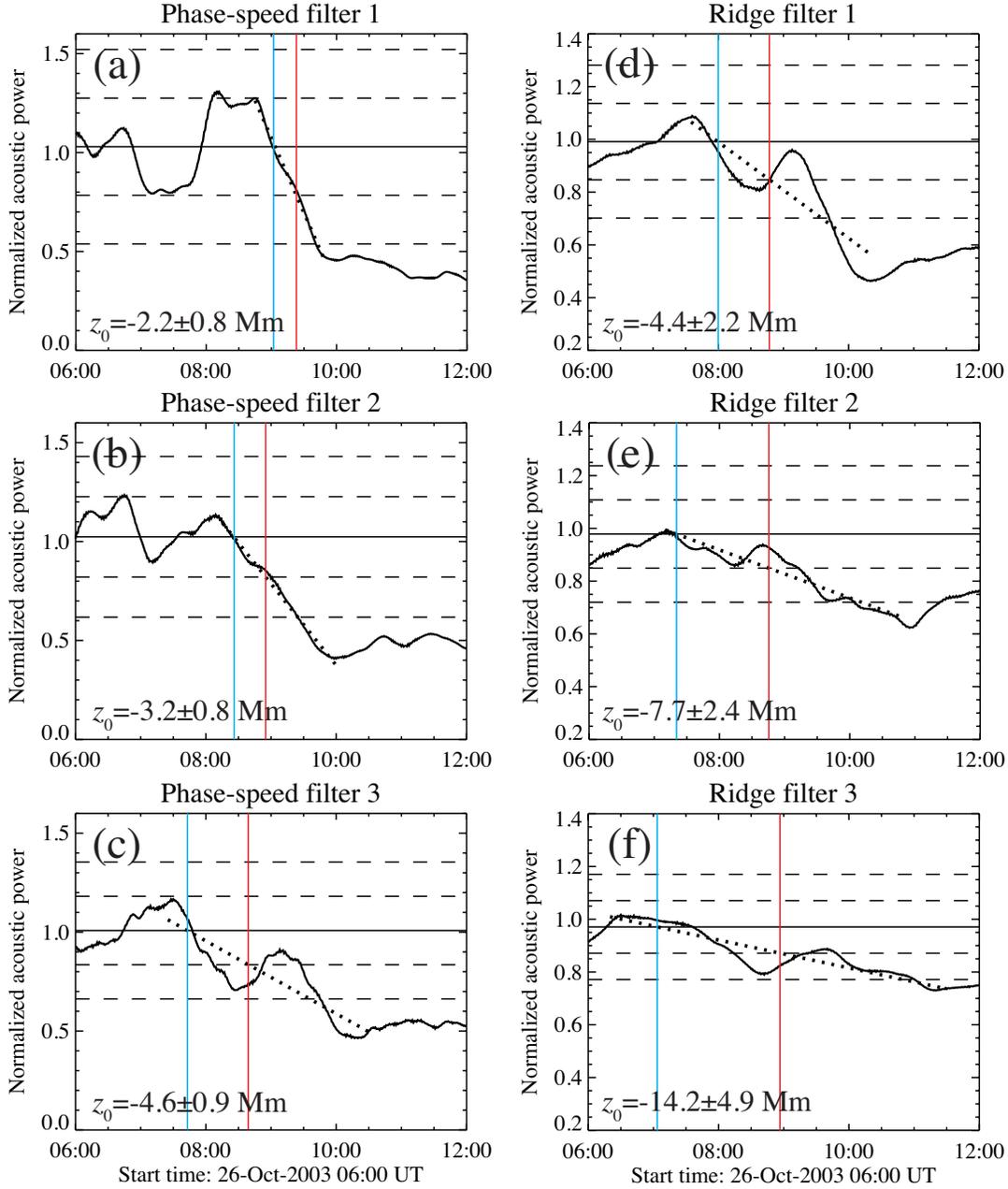}
  \caption{Acoustic power
    of the emerging AR 10488
    normalized by the quiet-Sun power
    for (a)--(c) phase-speed filters
    and (d)--(f) ridge filters.
    The horizontal lines
    (solid and dashed)
    are the mean,
    $\pm 1\sigma$,
    and $\pm 2\sigma$ power levels
    calculated from the surrounding region data.
    The dotted line is a fitted linear trend
    representing the power reduction,
    while blue and red vertical lines
    are the ``mean-crossing''
    and ``$-1\sigma$-crossing'' times
    of the dotted line,
    respectively.
    The target depth $z_{0}$
    is indicated
    in the bottom left
    of each panel.
  }
  \label{fig:filter}
\end{figure}

Figure \ref{fig:filter}
shows the temporal evolutions
of acoustic power
that correspond to
six different depth ranges
by phase-speed filters (a--c)
and ridge filters (d--f).
In each panel,
the plotted curve
is the acoustic power
from AR data
at the location
of flux emergence
$\mbox{\boldmath $x$}_{1}$,
normalized by the power
at the same location
but from QS data:
$P_{\rm AR} (\mbox{\boldmath $x$}_{1},t)/
P_{\rm QS} (\mbox{\boldmath $x$}_{1},t)$.
We apply 60-min
running average
($\pm$30-min from the target time)
to smooth measurements,
both in AR and QS,
to reduce rapid fluctuations
that may not correspond
to the subsurface magnetic field.
The horizontal lines are
the mean, $\pm 1\sigma$,
and $\pm 2\sigma$ levels
calculated from the quiet regions
surrounding the emerging AR.
Here, one can see that
the acoustic powers,
which are more or less unity
before 08:00 UT,
fall below $-2\sigma$ level
around 10:00 UT.
Considering that the flux
of AR 10488
appears at around 09:20 UT
by the method
introduced in \citet{tor12b},
the power reduction
seems to be highly related
to the magnetic field
of this emerging AR.
The amount of the reduction
is basically larger
for shallower filters.
The shallowest filter,
in Panel (a),
reveals the reduction up to 65\%.

To see the timing
of the power reduction
in Figure \ref{fig:filter},
we fit a linear trend to the curve
and measure 
the ``mean-crossing''
(reduction start)
and ``$-1\sigma$-crossing''
(significant reduction)
times.
The start of the fitting interval
is the last peak
greater than the mean level
that comes before the reduction slope
and the end is the point
where the slope becomes flattened
below the $-2\sigma$ level.
It is easily seen that
the mean-crossing times
are before 09:00 UT,
namely, before the flux appearance
at the photosphere,
and that
the mean-crossing time
becomes earlier
with depth.
Here, the deepest filter,
in Panel (f),
shows the earliest reduction,
which is more than 2 hr
before the flux appearance.

Figure \ref{fig:comparison}
shows the depth-time evolution of
the ``mean-crossing'' and ``$-1\sigma$-crossing''
of each filter
for (a) phase-speed filter,
(b) ridge filter,
and (c) both.
Here, the mean-crossings
in Panels (a) and (b)
show upward trends
from deeper to shallower
with time.
In this figure,
we also plot
the occurrence of the horizontal divergent flow (HDF),
which is the manifestation of the plasma
escaping from the rising magnetic field,
and the flux emergence
at the surface,
using the method
developed by \citet{tor12b}.
Therefore,
this figure indicates that
the rising patterns come
before the flux emergence,
or even before the HDF appearance
at the visible surface.
The mean-crossing
of the ridge filters
($-14$ to $-4.4$ Mm)
show a fast rising pattern
first at the rate of
several ${\rm km\ s}^{-1}$,
then at $\sim 1.5\ {\rm km\ s}^{-1}$,
while that
of the phase-speed filters
($-4.6$ to $-2.2$ Mm)
show the slower rate of
$\sim 0.5\ {\rm km\ s}^{-1}$.
In Figure \ref{fig:comparison}(c),
one can clearly see
the decelerating trend
of the mean-crossing times,
which will be discussed in detail
in the next section.
We confirmed that
even if we expand
the fitting intervals
in Figure \ref{fig:filter}
by 40 minutes,
at the cost of increased degree of misfit,
the mean- and $-1\sigma$-crossing times
shift by less than 15 minutes
(within the horizontal error bars
in Figure \ref{fig:comparison}c)
and thus the rising speed does not change much.

\begin{figure}
  \includegraphics[scale=1.,clip]{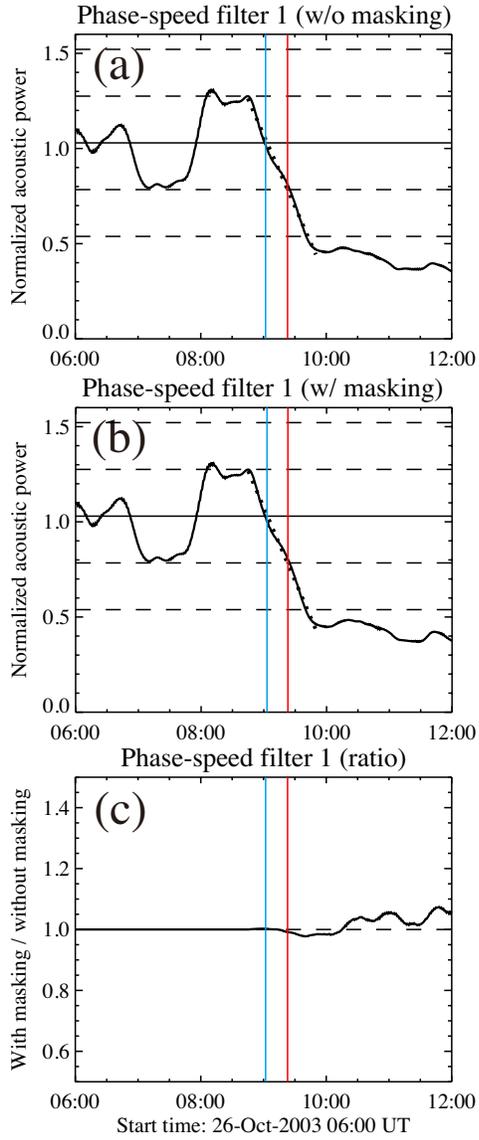}
  \caption{Normalized acoustic power
    for the shallowest filter,
    calculated (a) without masking
    and (b) with masking
    the pixels with field strength
    greater than 100 G
    in the averaging annulus.
    (c) The ratio of the power (b)
    over (a).
  }
  \label{fig:nomag}
\end{figure}

It is known that
the acoustic power is suppressed
in the surface magnetic fields.
Vertical fields may cause
mode conversion
of the acoustic waves
into down-going slow mode waves,
leading to
power reduction.
Other reduction
mechanisms include
energy conversion into thermal energy,
enhanced leakage
to atmosphere
due to changes in cut-off frequency,
emissivity reduction,
local suppression, etc. \citep[see][]{cho09}.
Numerous observations and theoretical works
have widely been carried out
in this field
(e.g., \citealt{bra87} for sunspots,
\citealt{jai09} for plage regions,
and \citealt{chi12} for small magnetic elements
in quiet region).
In order to examine
the effect of the surface field,
at least the direct and local effects,
we compare the acoustic power
calculated with and without masking
the strong surface field
in the averaging annulus.
Figure \ref{fig:nomag}(a)
shows the temporal evolution
of the normalized acoustic power
for the shallowest filter
(here we call ``without masking''),
which is the same as Figure \ref{fig:filter}(a).
To reduce the effect of the field,
we also calculated the power
by excluding the pixels with field strength
greater than $100$ G
from the averaging annulus
(``with masking''),
which is shown as
Figure \ref{fig:nomag}(b).
Here, the fitted (dotted) line
is found to be
just the same as that in Panel (a),
and thus the mean-crossing and $-1\sigma$-crossing
times do not change.
Panel (c) shows the ratio of the power
(b) with masking over (a) without masking.
Here,
the ratio is deviated from unity
in the later time,
which indicates that
in fact we see
the effect of the surface field.
However,
when the flux first appears at 09:20 UT,
the difference is less than 5\%,
which increases afterward
but remains less than 10\%.
It is because
the fraction of magnetized pixels
to the total number of pixels
in the annulus
is small (of the order of a few percent).
As for the other five deeper filters,
the difference is much smaller
and the mean- and $-1\sigma$-crossing times
do not change at all.
Therefore,
we can conclude that,
although the emergence starts at 09:20 UT
and the temporal power-smoothing
has a $\pm$30-min window from the target time,
the effect of the surface field
on the power reduction
at around and before 08:50 UT
is fairly negligible.
Note that
this masking method
does not remove
the surface field effect perfectly,
since weaker pixels
(field strength $\le 100\ {\rm G}$)
may also be affected by the surrounding fields
and unresolved kG strength flux tubes
may exist in such pixels.
Thus,
wave absorption
or mode conversion
by surface fields
(maybe associated with the rising flux)
may play a role in the observed acoustic power reduction,
particularly after the emergence of the flux.
To measure to which extent
the surface field affects
our measurements,
we need to repeat
our analysis on
many more regions,
e.g.,
regions with similar flux distributions
but without emerging ARs,
or with other ARs.

\section{Discussion and Conclusions
  \label{sec:discussion}}

\begin{figure}
  \includegraphics[scale=1.,clip]{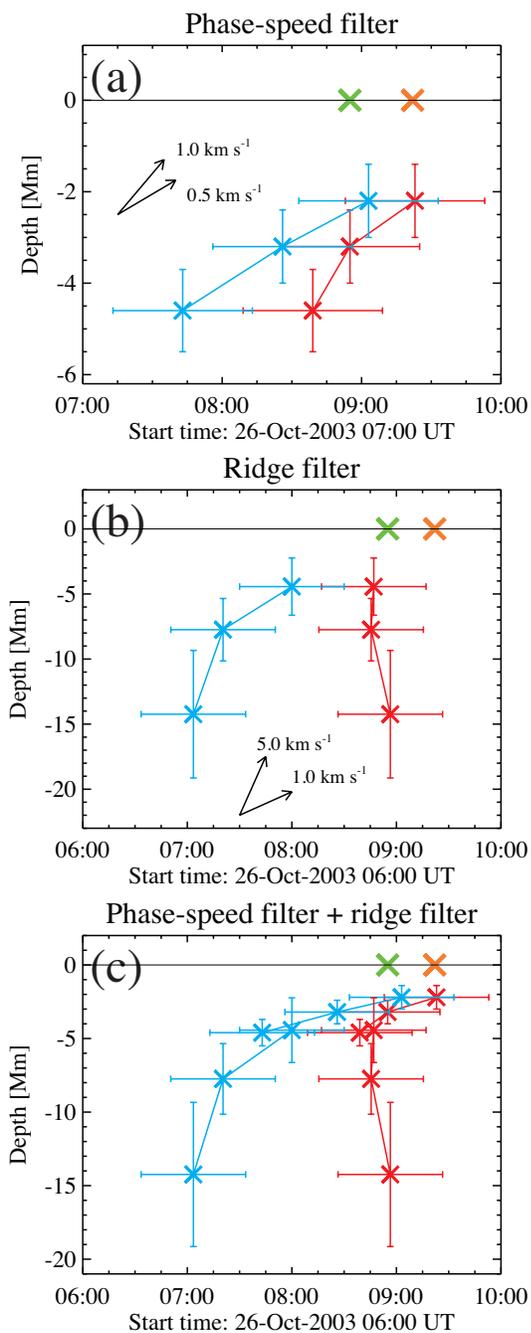}
  \caption{
    Depth-time evolution
    of the ``mean-crossing''
    (reduction start; blue)
    and ``$-1\sigma$-crossing''
    (significant reduction; red)
    for (a) phase-speed filters,
    (b) ridge filters,
    and (c) both.
    The vertical and horizontal error bars indicate
    the uncertainty in the effective target depth
    of each filter
    and 60-min smoothing average,
    respectively.
    Green and orange X's
    are the occurrence time
    of horizontal divergence flow (HDF)
    and the flux appearance
    measured by the method
    in \citet{tor12b}.}
  \label{fig:comparison}
\end{figure}

As is evident
in Figure \ref{fig:comparison},
before the emerging flux appears
at the visible surface,
the onset of the acoustic power reduction
starts from the deeper layers,
and the speed of the rising trend
gradually changes
from several to less than $1\ {\rm km\ s}^{-1}$
in the shallower convection zone
($>-20\ {\rm Mm}$),
suggesting the deceleration
of the power-reducing agent.
If we assume that
this is actually a magnetic field,
Figure \ref{fig:comparison}
means that the magnetic flux
shows the rising motion,
first at the rate of
$6\ {\rm km\ s}^{-1}$
in the $15$--$10$ Mm depth range,
then at the rate of
$1.5\ {\rm km\ s}^{-1}$
in $10$--$5$ Mm,
finally at $0.5\ {\rm km\ s}^{-1}$
in $5$--$2$ Mm.
This gradual deceleration
of the emerging magnetic flux
is well in line with
the theoretical ``two-step emergence'' model
by \citet{tor10}.
In this model,
the emerging flux
in the uppermost convection zone
is decelerated
because of the photospheric layer
ahead of the flux,
which then triggers
the magnetic buoyancy instability
to penetrate the photosphere,
and eventually restart rising
into the solar atmosphere,
leaving an HDF just before
the flux appearance
at the visible surface.
The deceleration 
may be more effective
in the shallower layer
above $-20$ Mm,
since, at around $-20$ Mm,
the radius of the rising flux tube
exceeds the local pressure scale height \citep{fan09},
which encourages the mass accumulation
and the resultant deceleration.
By considering this model,
we can speculate that
the deceleration
in the shallower layer
and the flux appearance
at the surface
shown in Figure \ref{fig:comparison}
are the manifestation
of the two-step emergence model.

In \citet{ilo11},
the average emergence speed
of the magnetic flux
in AR 10488
from $-65$ Mm to the surface
is estimated to be
$0.6\ {\rm km\ s}^{-1}$,
while,
in \citet{ilo12},
the speed from $-70$ to $-50$ Mm
is estimated to be
$\sim 1\ {\rm km\ s}^{-1}$.
Also, from the thin-flux-tube simulation,
the rising speed is about $1\ {\rm km\ s}^{-1}$
at $-10$ Mm \citep{fan09}.
In the present study,
the rising speed between
$0.5$ and $1.5\ {\rm km\ s}^{-1}$
(namely, of the order of $1\ {\rm km\ s}^{-1}$),
measured from the five filters
in the upper $10\ {\rm Mm}$,
are consistent with
previous studies.
In addition,
we find that
the flux went through
in the upper $\sim 15\ {\rm Mm}$
within $\sim 2$ hours,
which is also in agreement with
previous helioseismic studies
\citep{kos09}.

One important factor
we should take into account
is the difference of the sensitivity
to the power reduction
between the two types of the filters
(phase-speed filters and ridge filters),
or even among the filters
of the same group
but for different depths.
Here we expect that
the measurements of
the power-reduction
due to rising magnetic fields
will be less sensitive
when the fields are
at larger depths,
since (1) large depths
are probed
by acoustic waves
with large horizontal wavelengths
and, for large wavelengths,
the absorption may be less effective,
and (2) the conversion
of acoustic waves into slow MHD waves,
one of the main power-reduction mechanisms
\citep{cal03},
is probably less effective
at large depths,
where the gas pressure
dominates over
the magnetic pressure.
It was also shown that,
in the case of sunspot fields,
the absorption coefficient drops
to almost zero
at depths of about $15$--$20\ {\rm Mm}$
\citep{ilo11b},
which is the target depth
of the deepest filter.
Here the rising speed
is simply calculated from
the difference of target depths
of two filters
over the detection time difference.
If the deepest filter
is less sensitive,
the detection time
by this filter
might be later
than the actual time
and thus the rising speed
might be overestimated
at $6\ {\rm km\ s^{-1}}$.

These uncertainties
make it difficult
to directly compare
the power reduction events
in Figure \ref{fig:comparison}.
Therefore,
a clear identification
of the power-reducing agent
requires much work,
which we shall leave for future research.
Nevertheless,
we find a rising motion
which is related to the flux emergence,
prior to the flux appearance
at the photosphere.

In this Letter
we apply a set of phase-speed filters
and ridge filters
to the {\it SOHO}/MDI Dopplergrams
of the emerging AR 10488
to see the acoustic power reduction
at different depths.
In summary,
our results show the following:
\begin{enumerate}
  \item All of the investigated acoustic powers
    show reductions,
    up to more than 2 hr
    before the flux appearance
    at the photosphere.
  \item In both filter groups,
    the start times of the power reduction
    show a rising trend
    and a gradual deceleration.
    The trend speed is
    first $6\ {\rm km\ s}^{-1}$
    in the 15--10 Mm depth range,
    then $1.5\ {\rm km\ s}^{-1}$
    in 10--5 Mm,
    finally $0.5\ {\rm km\ s}^{-1}$
    in 5--2 Mm.
  \item If we assume that
    the power reduction
    is caused by a magnetic field
    corresponding to AR 10488,
    the detected deceleration
    is well in accordance with
    the two-step emergence model
    of the emerging magnetic field.
    This study observationally
    supports
    the theoretical two-step model.
  \item The estimated emerging speeds
    of about $1\ {\rm km\ s}^{-1}$
    are highly consistent
    with other observations
    and numerical simulations.
    The speed
    at larger depths,
    however,
    may be overestimated
    with this method.
    We should examine and improve
    the present analysis method
    and investigate
    how sensitive each filter is
    to the target region,
    and what the detected object actually is.
\end{enumerate}

Although this work
shows a promising result,
here we just analyzed
one particular set
of AR and QS.
Therefore,
we need to repeat our measurements
on many more regions.

\acknowledgments

The authors appreciate the useful comments
by the anonymous referee.
S.T. would like to thank
T. Hoeksema and K. Hayashi
for inviting to visit
Stanford University.
S.T. is supported
by the JSPS Institutional Program
for Young Researcher Overseas Visits,
and by the Grant-in-Aid for JSPS Fellows.

%% The reference list follows the main body and any appendices.
%% Use LaTeX's thebibliography environment to mark up your reference list.
%% Note \begin{thebibliography} is followed by an empty set of
%% curly braces.  If you forget this, LaTeX will generate the error
%% "Perhaps a missing \item?".
%%
%% thebibliography produces citations in the text using \bibitem-\cite
%% cross-referencing. Each reference is preceded by a
%% \bibitem command that defines in curly braces the KEY that corresponds
%% to the KEY in the \cite commands (see the first section above).
%% Make sure that you provide a unique KEY for every \bibitem or else the
%% paper will not LaTeX. The square brackets should contain
%% the citation text that LaTeX will insert in
%% place of the \cite commands.

%% We have used macros to produce journal name abbreviations.
%% AASTeX provides a number of these for the more frequently-cited journals.
%% See the Author Guide for a list of them.

%% Note that the style of the \bibitem labels (in []) is slightly
%% different from previous examples.  The natbib system solves a host
%% of citation expression problems, but it is necessary to clearly
%% delimit the year from the author name used in the citation.
%% See the natbib documentation for more details and options.

\clearpage

\clearpage

\clearpage

\clearpage

\clearpage

\begin{deluxetable}{cccc}
  \tablewidth{0pt}
  \tablecaption{Filtering parameters
    used in this Letter
    \label{tab:filter}}
  \tablehead{
    \colhead{Filter \#} &
    \colhead{Target depth $z_{0}$} &
    \colhead{Phase speed $V_{\rm ph}$} &
    \colhead{Annulus range $\Delta$} \\
    \colhead{} &
    \colhead{[Mm]} &
    \colhead{[${\rm km\ s}^{-1}$]} &
    \colhead{[Mm]}
  }
  \startdata
   Phase-speed filter\tablenotemark{a}: & & & \\
   1 & $-2.2\pm 0.8$ & $14.9\pm 2.2$ & 6.6--9.5 \\
   2 & $-3.2\pm 0.8$ & $17.6\pm 2.2$ & 9.5--12.4 \\
   3 & $-4.6\pm 0.9$ & $21.5\pm 2.7$ & 13.1--16.0 \\
   \hline
   Ridge filter\tablenotemark{b}: & & & \\
   1 & $-4.4\pm 2.2$ & $21.0\pm 6.0$ & 7.1--15.5 \\
   2 & $-7.7\pm 2.4$ & $31.0\pm 7.0$ & 13.7--24.3 \\
   3 & $-14.2\pm 4.9$ & $46.0\pm 10.0$ & 22.0--46.2 \\
  \enddata
  \tablenotetext{a}{
    The phase speed
    and annulus range
    are cited from \citet{zha12},
    while the target depth is calculated
    from the model S of \citet{chr96}.}
  \tablenotetext{b}{After the filtering,
    the effective phase speed
    is evaluated,
    then the depth and annulus range
    are calculated from the model S.}
\end{deluxetable}

\end{document}